\documentclass[twocolumn,showpacs,preprintnumbers,amsmath,amssymb,prb,superscriptaddress]{revtex4}

\usepackage{graphicx}
\usepackage{color}

\newfont{\wasy}{wasy10 scaled 1000}

\newcommand{\n}{n}

\newcommand{\Mat}{{\cal S}}
\newcommand{\lesser}{{-+}}
\newcommand{\greater}{{+-}}

\newcommand{\nt}{\tilde{n}}

\begin{document}

\title{Anderson impurity model in nonequilibrium: analytical results
versus quantum Monte Carlo data }

\author{L. M\"uhlbacher}
\affiliation{Physikalisches Institut, Albert-Ludwigs-Universit\"at
Freiburg, Hermann-Herder-Str. 3, D-79104 Freiburg, Germany}

\author{D. F. Urban}
\affiliation{Physikalisches Institut, Albert-Ludwigs-Universit\"at
Freiburg, Hermann-Herder-Str. 3, D-79104 Freiburg, Germany}

\author{A. Komnik}
\affiliation{Institut f\"ur Theoretische Physik,
Ruprecht-Karls-Universit\"at Heidelberg,\\
 Philosophenweg 19, D-69120 Heidelberg, Germany}

\date{\today}

\begin{abstract}
  We analyze the spectral function of the single-impurity two-terminal Anderson
  model at finite voltage using the recently developed
  diagrammatic quantum Monte Carlo technique as well as perturbation theory.
  In the (particle-hole-)symmetric case we find an excellent agreement of the
  numerical data with the perturbative results of second order up to interaction
  strengths $U/\Gamma \approx 2$, where $\Gamma$ is the transparency of the
  impurity-electrode interface. The analytical results are obtained in form of the
  nonequilibrium self-energy for which we present explicit formulas
  in the closed form at arbitrary bias voltage. We observe an increase of the spectral
  density around zero energy brought about by the Kondo effect. Our analysis
  suggests that a finite applied voltage $V$ acts as an
  effective temperature of the system. We conclude that at voltages significantly
  larger than the equilibrium Kondo temperature there is a complete suppression of
  the Kondo effect and no resonance splitting can be observed. We confirm this scenario by
  comparison of the numerical data with the perturbative results.
\end{abstract}


\maketitle

\section{Introduction}
Despite its simplicity the Anderson impurity model
(AIM)\cite{Anderson1961} contains a surprising amount of
interesting physics. Under equilibrium conditions, when the
singly occupied dot level lies deep below the chemical potential
of the fermion continuum and when the Coulomb repulsion $U$
prohibits double occupation of the dot, the impurity spectral
function, i.e.\ the impurity density of states (DOS), is known to
develop a sharp Kondo (aka Abrikosov-Suhl) resonance which is
located at the chemical potential of the fermion continuum. It is
observable at relatively low temperatures $T < T_K$, where the
Kondo temperature $T_K$ is an estimate for the resonance width. This is the essence of the conventional Kondo effect.\cite{Hewson1997,Tsvelick1983}

In recent years a quite natural extension of
Anderson's original idea, namely a setup in which the impurity is
coupled to two fermion continua,  came to attention as its
experimental realization became feasible.
\cite{Goldhaber-Gordon1998,Cronenwett1998,Schmid1998} A
particularly interesting direction of research is the
investigation of the nonequilibrium transport as well as of the Kondo effect
which has been
quite successfully approached analytically, see e.~g. Refs.
[\onlinecite{Hershfield1992,Meir1993,Haule2001,Oguri2002,Konik2002,Hettler1998,
Komnik2004,Ratiani2009,Schoeller2009,Korb2007}], as well as
numerically, see e.~g. Refs.
[\onlinecite{Costi1993,Anders2008,Kirino2008,Heidrich-Meisner2009,
Rincon2009,Gezzi2007}].  Yet another but related research direction is concerned with the nonequilibrium transport in a pure Kondo model, see e.~g. Refs.~[\onlinecite{Schiller1998,Coleman2001,Rosch2001,Kehrein2005,Pletyukhov2010}].

In order to induce a finite electric current through a
quantum dot at least two electrodes with \emph{different} chemical
potentials are necessary. Contrary to the naive expectation that
coupling to two fermionic continua may lead to a two-channel Kondo
effect the original resonance was predicted to split into two
peaks which are positioned at the two chemical
potentials.\cite{Coleman2001,Pustilnik2004,Meir1993,Wingreen1994,Rosch2001} Electron
transport between the leads is accompanied by spin-flip processes
which essentially break the symmetry necessary to drive the system
into the two-channel Kondo fixed point. On the other hand, exactly
those spin-flip processes participate in the RG flow and grow
stronger toward low energies thereby being responsible for almost
perfect effective electron transmission (at the Fermi edge in
equilibrium) across the impurity far below $T_K$. In the case of a
finite applied voltage $V$ there is a constant spin-flip rate
associated with the current flowing through the system, so that
the Kondo peaks are weaker and broadened in comparison to the
equilibrium situation.\cite{Meir1993,Kaminski2000} This splitting
as well as the broadening of the Kondo peaks are difficult to
access in the two-terminal setup. Therefore a three-terminal
approach for the spectral function measurement has been put
forward,\cite{Lebanon2001,Sun2001} which shortly afterwards was
implemented experimentally.\cite{DeFranceschi2002,Leturcq2005}

Although by now a number of studies of multi-terminal
Kondo/Anderson setups have been conducted, see e.~g.\ Refs.\
[\onlinecite{Shah2006,HBTLetter}], it is desirable to analyze the
problem with non-approximative methods in order to cover the full
Kondo crossover.\cite{Costi1993} The recently developed diagrammatic
quantum Monte Carlo (diagMC) approach not only allows to simulate
finite voltage transport but also reliably works even at zero
temperature.\cite{Muhlbacher2008,Schmidt2008,Werner2009,schiro_PRB_79,Werner2010}
In this paper the diagMC is applied in order to calculate the
impurity spectral function of the AIM under nonequilibrium
conditions with special emphasis on the Kondo effect related
features. The numerical data is then compared with the results of
the analytical perturbative expansion in interaction strength $U$.
We find that as soon as the bias voltage exceeds the equilibrium
$T_K$ the Kondo effect related features in the spectral function
rapidly deteriorate. By comparison of the numerical data with the
perturbative results we conclude that the large finite voltage has
similar consequences as finite temperature. This explains why the
Kondo features are so weak and the resonance doubling is not
observable at all.

The outline of the paper is as follows: In Sec.\ \ref{Sec:PertTheory}
we formulate the problem and identify the most interesting energy
regimes. The perturbative results for the self-energy and impurity DOS,
for which we obtain analytical expressions in a closed form, are presented.
Section \ref{Sec:DQMC} is devoted to the details of the numerical implementation
of the diagMC scheme. Results are presented in Section \ref{Sec:Discussion}, where
the numerical data are compared with the perturbative result and the
physical picture is discussed.

\begin{figure}[t]
\begin{center}
    \includegraphics[width=\columnwidth]{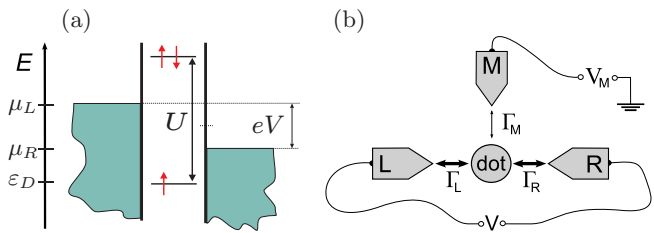}
    \caption{(a) Level structure of the quantum dot
    with respect to the two fermion continua (leads) at chemical potentials $\mu_{L/R}$.
    The dot level $\varepsilon_D$ corresponding to single occupation lies below $\mu_{L/R}$
    while the energy level for double occupation $\varepsilon_D+U>\mu_{L/R}$ is augmented by
    $U$ due to Coulomb repulsion. (b) Sketch of the setup. For the analytical analysis we
    consider the two terminal setup in which the quantum dot is coupled to a left (L) and
    right (R) lead via electron tunneling $\propto\Gamma_{L/R}\equiv\Gamma$. For the numerical
    analysis it is advantageous to add a third measuring electrode $M$ which is only very
    weakly coupled to the quantum dot´ and to consider the limit $\Gamma_M\rightarrow 0$.
    \label{fig:setup}}
\end{center}
\end{figure}

\section{Formulation of the problem and perturbative results}
\label{Sec:PertTheory}
We model the system in the canonical way using
the Hamiltonian
\begin{eqnarray}
\label{eq:H}
    H = H_0 + H_I + H_T \, .
\end{eqnarray}
Here $H_0$ contains both the uncoupled impurity level at energy $\epsilon_D$ which can
be subject to a (local) magnetic field $h$, as well as the left/right
contacting leads. The latter are assumed to be noninteracting fermion continua
with field operators $\psi_{\alpha,\sigma}$, $\alpha=L/R$, kept at chemical
potentials $\mu_{L/R}$. Here $\sigma$ is the spin index.
\begin{eqnarray}
\label{eq:H0}
   H_0 =\sum_{\alpha,\sigma}  H_\alpha[\psi_{\alpha,\sigma}] + \sum_\sigma
   (\epsilon_D + \mu_B g \sigma h/2)
   \, d^\dag_{\sigma} d_{\sigma} \,,
\end{eqnarray}
where $d_\sigma$ and $d^\dag_\sigma$ are the annihilation/creation
operators of an electron on the impurity level. As usual, $\mu_B$
is Bohr's magneton and $g$ is the Land\'{e} factor. Electron
exchange between the electrodes and the impurity takes place
locally at $x=0$ and is accomplished by
\begin{eqnarray}
   H_T = \gamma \sum_{\alpha, \sigma}
    \, d_{
   \sigma}^\dag \psi_{\alpha, \sigma}(0) + \mbox{H.c.} \, ,
\end{eqnarray}
where $\gamma$ is the tunneling amplitude between dot
level and electrode, which we for simplicity assume to be equal for both
contacts.\footnote{A generalization to an asymmetrically coupled system is straightforward. To simplify notation we concentrate on the symmetric system.} Finally, Coulomb repulsion on the
impurity is taken into account via the last term,
\begin{eqnarray}
   H_I = U \,  n_{\uparrow} \, n_{\downarrow} \, ,
\end{eqnarray}
where $n_{\sigma} = d^\dag_{\sigma}  d_{\sigma}$. Perhaps the most
interesting parameter regime is the particle-hole symmetric one,
when $h=0$ and $\epsilon_D = - U/2$ (aka \emph{symmetric} Anderson
model), for which the level structure of the quantum dot is depicted in
Fig.\ \ref{fig:setup}(a).
Under equilibrium conditions this model is solvable by the
Bethe Ansatz.\cite{Tsvelick1983,Andrei1983} This method works
perfectly as far as thermodynamic properties are concerned but the
extraction of single particle quantities, although in principle
possible, still remains an open issue. The most important single
particle quantity is the local impurity DOS (spectral function)
$\rho_d(\omega)$. It is related to the Fourier transform of the
retarded Green's function (GF) of the dot $D^R(\omega)$ via
\begin{eqnarray}
\label{eq:rhodefinition}
   \rho_d(\omega) = - 2 \mbox{Im} \, D^R(\omega) \, ,
\end{eqnarray}
where
\begin{eqnarray}
\label{eq:DRdefinition}
   D^R(t) = - i \Theta(t) \langle \left\{ d_\sigma(t), \,
   d_\sigma^\dag(0) \right\}
   \rangle \, .
\end{eqnarray}
Here $\{.,.\}$ denotes the anticommutator and $\Theta(t)$ is the
Heavyside step function. In the interacting case,  $D^R(\omega)$
can be expressed in terms of the self-energy $\Sigma^R(\omega)$,
\begin{eqnarray}
\label{eq:DRdefinition:B}
    D^R(\omega)&=&\frac{1}{\omega-i\Gamma-\Sigma^R(\omega)}.
\end{eqnarray}
Here $\Gamma = 2 \pi \rho_0 |\gamma|^2$ is the lead-dot contact
transparency that depends on the tunneling amplitude
$\gamma$ and on the local DOS in the electrodes $\rho_0$, which
we assume to be only weakly energy-dependent.
For simplicity of notation, we will use units for which $\Gamma\equiv 1$
in the following, so that $\Gamma$ is the unit of energy.

As has been pointed out in Ref.\ [\onlinecite{Meir1992}] the spectral
function plays an important role especially in nonequilibrium
transport and the current through the dot is
given by the Meir-Wingreen formula\cite{Meir1992}
\begin{eqnarray}
    I(V) = G_0 \int d \omega \, \rho_d(\omega) \, \left[n_L(\omega) - n_R(\omega)\right] \, ,
\end{eqnarray}
where $G_0=2 e^2/h$ is the conductance quantum and $n_{L,R}
(\omega) = \left[ \exp\left((\omega - \mu_{L,R})/T\right) +
1\right]^{-1}$ are the Fermi distribution functions in the
respective electrodes. First calculations of the spectral function
use perturbation theory in
$U$.\cite{Yamada1975,Yamada1975a,Yosida1975,Yosida1970,
Zlati'c1980,Zlatiifmmodeacutecelse'cfi1983} The corresponding
series turns out to be well-controlled and rapidly converging.
Often there is no necessity for having the complete energy
dependence of the spectral function (via e.g.\ the corresponding
self-energies) at hand so that an expansion of $\rho_d(\omega)$
for low energies is sufficient. In all other situations there
exist numerous studies also in
nonequilibrium.\cite{Hershfield1992,Oguri2002,M.Hamasaki2006,Muhlbacher2008,
Schmidt2008,Weiss2008,Werner2009,schiro_PRB_79,Werner2010,MuehlbacherUnpublished,Fujii2003}
However, analytical expressions in a closed form for the
self-energy from the second order onwards do not yet exist.

For technical reasons the calculation of the self-energies has a
number of advantages. Starting point are
the impurity GF without interactions.
With $H_I\equiv 0$ the Hamiltonian (\ref{eq:H}) is quadratic and can be
trivially diagonalized even in nonequilibrium.\cite{Caroli1971}
The noninteracting GF (we concentrate on the electron-hole symmetric case)
read\cite{AndersonFCS,Egger2008}
\begin{eqnarray}
\label{eq:RAK}
   D^{R,A}(\omega) &=& \frac{1}{\omega \pm i} \, ,
   \nonumber \\
   D^K(\omega) &=& \frac{2i(n_L(\omega) + n_R(\omega) - 1)}{\omega^2+1} \, .
\end{eqnarray}
Just as well we can work with the Keldysh GF $D^{kl}(\omega)$ which
are related to the GFs in the RAK representation (\ref{eq:RAK}) by a simple rotation in
Keldysh space.\cite{Lifshits1981}
As usual, the Keldysh indices $k,l=\pm$ denote the branch of the
contour on which the final and initial times of
the corresponding GF in the time domain are taken: $-$ for the time-ordered one and
$+$ for the anti-time-ordered one.

\begin{figure}[]
\begin{center}
    \includegraphics[width=7cm]{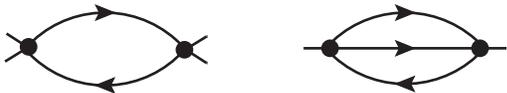}
    \caption{Feynman diagrams for the polarization loop (left), and
    for the self-energy at second order, $\propto U^2$  (right).
    \label{fig:diagrams}}
\end{center}
\end{figure}

\begin{figure}
\begin{center}
   \includegraphics[width=\columnwidth]{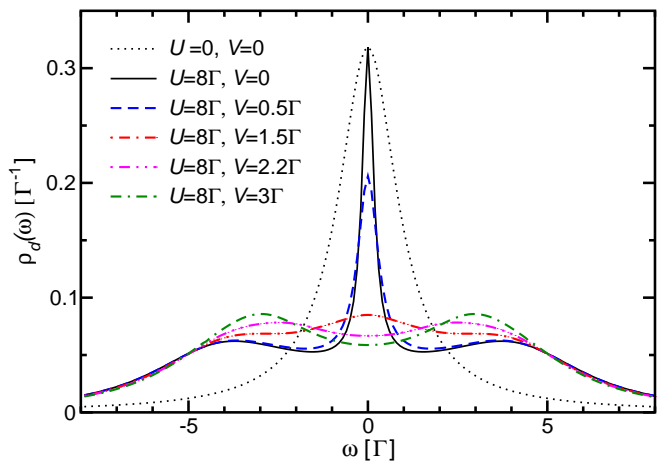}
   \caption{Zero temperature dot spectral function for different
   interaction strengths and bias voltages. The curves shown are deduced
   from the analytical result (\ref{eq:SigmaResult}) of the perturbative calculation.
   \label{fig:Results:PertTheory}}
\end{center}
\end{figure}

The lowest order self-energy, c.f.\ Fig.\ \ref{fig:diagrams}, contains two energy integrations
and is given by
\begin{eqnarray}
    \Sigma^{kl}(\omega) &=& - \int \frac{d \Omega \, d \epsilon}{(2\pi)^2}
    \, D^{kl}(\omega - \Omega) \, D^{lk}(\epsilon) \,
    D^{kl}(\epsilon + \Omega)
\nonumber \\
    &=&  i \int \frac{d \epsilon}{2 \pi} \, D^{kl}(\omega + \epsilon) \,
    \Pi^{lk}(\epsilon)  \, ,
    \label{eq:SigmaIntegral}
\end{eqnarray}
where
\begin{eqnarray}
    \Pi^{lk}(\omega) = i \int \frac{d \epsilon}{2 \pi} \, D^{lk}
    (\omega + \epsilon) \, D^{k l}(\epsilon) \,
\end{eqnarray}
is the generalized (Keldysh) polarization loop.\footnote{The retarded polarization loop $\Pi^R(\omega)$ is directly related to the
dynamical spin/charge susceptibilities.\cite{Yosida1975,Zlatiifmmodeacutecelse'cfi1983}}
An explicit calculation of these quantities is tedious but straightforward. In
the following we give the results only. The retarded polarization
loop is given by\cite{Egger2008}
\begin{eqnarray}
\label{eq:def:PiR}
    \Pi^R(\omega) &=& \Pi^{--}(\omega) - \Pi^{-+}(\omega)
    \\
    &=& - \frac{1}{\pi} \frac{1}{\omega^2 + 2i\omega} \ln
    \left[ \frac{(1 - i \omega)^2 + (V/2)^2}{1 + (V/2)^2 } \right]
\end{eqnarray}
and we find the time-ordered polarisation loop to read
\begin{eqnarray}
    &&\Pi^{--}(\omega)=\Pi^R(|\omega|)\quad
\nonumber\\
    &+&\frac{i}{\pi}
    {\rm Im}
    \!\left[\frac{1}{\omega^2\!+\!2i|\omega|}\ln\!\left(\!\frac{|\omega|-V/2+i}{V/2+i}\!\right)\!\right]
    \Theta(|V|-|\omega|).\qquad
\end{eqnarray}
The remaining Keldysh components can be inferred from relation
(\ref{eq:def:PiR}) and the symmetry properties of $\Pi^{kl}$,
namely $\Pi^{+-}(\omega)=\Pi^{-+}(-\omega)$ and
$\Pi^{++}(\omega)=-\Pi^{--}(\omega)^*$. With these expressions at
hand we perform the integral (\ref{eq:SigmaIntegral}) and find the
retarded self-energy to be given by
\begin{widetext}
\begin{eqnarray}
\label{eq:SigmaResult}
    \Sigma^R(\omega) &=& -\frac{3i }{(2 \pi)^2} \left\{
    \frac{1}{1 + \omega^2} \left[\frac{\pi^2\omega}{3i}
    +\frac{1}{2} \ln^2 \left( \frac{i-V/2}{i+V/2} \right)
    +\sum_{\alpha,\beta=\pm} \mbox{Li}_2 \left(\beta\frac{\alpha V/2-\omega}{\alpha V/2+\beta i} \right)\right]
    \right.
\nonumber \\
    &&+ \left.
    \frac{1}{(i + \omega)(3 i + \omega)} \left[\pi^2
    -\frac{1}{2} \ln^2 \left( \frac{i-V/2}{i+V/2} \right)
    -\sum_{\alpha,\beta=\pm} \Lambda\left(\frac{2i+\omega-\alpha\beta V/2}{i+\alpha V/2}\right)\right]\right\}.
\end{eqnarray}
\end{widetext}
Here ${\rm Li}_2$ denotes the dilogarithm function and we have introduced
\begin{equation}
    \Lambda(z)=\mbox{Li}_2(z)-i\pi\mbox{sgn}[\mbox{Im}(z)]\ln(z).
\end{equation}
In equilibrium this expression simplifies considerably,
\begin{eqnarray}
    \Sigma_{eq}^R(\omega) &=& - \frac{1}{\pi^2 (1 + \omega^2)}
    \left( \frac{\pi^2}{4} \omega + 3 i \, \mbox{Li}_2(i \omega)
    \right)
\nonumber\\
    &&+\frac{3 i }{\pi^2(i+ \omega)(3 i + \omega)}\left[
    \Lambda(2 - i \omega) - \frac{\pi^2}{4} \right].\quad
\end{eqnarray}
Finally, the spectral function is given via Eqs.\ (\ref{eq:rhodefinition}, \ref{eq:DRdefinition:B}).
Its behavior as function of energy is depicted in Fig.\ \ref{fig:Results:PertTheory}
and discussed in section \ref{Sec:Discussion}.

\section{Diagrammatic Monte Carlo simulation method}
\label{Sec:DQMC} The recently developed refinements of the diagMC
technique allow to access regimes of arbitrary interaction
strength and therefore make diagMC simulations a suitable tool for
numerically investigating the AIM with large onsite Coulomb
repulsion\cite{Werner2010}.
Moreover, this approach gives access to the transient behavior of
the system after sudden switching of the tunneling coupling and
therefore contributes to the rapidly developing research area
investigating interaction quenches in quantum dots, see e.g.\ Refs.\
[\onlinecite{Anders2005,Komnik2009,Eckstein2010,Ratiani2010,Heyl2010,Pletyukhov2010,Karrasch2010}].
An advantage of diagMC is the possibility to explicitly implement
the band structure of the electrodes. In this regard it can be
seen as an ideal tool for discussing experimental results.

In principle, the dot spectral density can be measured using its immediate definition (\ref{eq:rhodefinition}).
However, as suggested in Refs.\ [\onlinecite{Lebanon2001,Sun2001}], the spectral
function can more efficiently be deduced by adding an additional third probe electrode,
which is only weakly coupled to the quantum dot, c.f.\ Fig.\ \ref{fig:setup}(b).
The conductance of this `measuring' electrode (denoted by the subscript `M' in our
notation) is given by
\begin{equation}
\label{eq:g_M 0}
    g_M(t) = \frac{d}{d\mu_M} I_M(t) \,,
\end{equation}
where $I_M$ and $\mu_M$ are the current and chemical potential for the third
contact, respectively. In the limit of vanishing coupling $\Gamma_M$,
the spectral density is related to the steady state value ($t\rightarrow\infty$)
of the conductance by\cite{Lebanon2001,Sun2001}
\begin{equation}
    \rho_d(\mu_M)=\lim_{\Gamma_M\rightarrow 0} \Gamma_M^{-1} \lim_{t\rightarrow\infty} g_M(t)\,.
\end{equation}
An advantage of this approach is that it corresponds to the
method used in experimental studies.\cite{Leturcq2005}

Following the lines of Refs.\
[\onlinecite{Muhlbacher2008,Werner2009,MuehlbacherUnpublished}]
the current through the contact $\alpha$, with
$\alpha\in\{L,\,M,\,R\}$, in the absence of magnetic fields is
given by
\begin{eqnarray}
\label{eq:I_alpha}
    I_{\alpha \sigma}(t)
    =
    - 4
    \text{Im} \sum_{k \in L} \gamma_{\sigma k}(t)\,
    \text{tr}\!\left\{W_0\, a_{\alpha\sigma k}^\dag(t)\, d_\sigma(t) \right\} \,,
\end{eqnarray}
where $a_{\alpha \sigma k}^\dag$ is the momentum-$k$ component of the
$\psi^\dag_{\alpha \sigma}(x)$ field operator and $W_0$ denotes
the full initial density operator of the system (which we assume
to factorize, with the dot being initially empty).
Note that the tunneling amplitude $\gamma_{\alpha k}$ in
Eq.\ (\ref{eq:I_alpha}) is explicitly energy/momentum dependent.
This is necessary because contrary to
the analytical calculations of the previous section, where an
infinite flat band significantly simplifies the calculations, the
use of a band with a finite width is essential in numerical
approaches for exactly the same reason. An additional speed-up of
the diagMC simulations is achieved by a smooth switch-on procedure
accomplished by a time dependent coupling
\begin{equation}
\label{t(t)}
    \gamma_{\alpha k}(t) = g_\alpha(t)\, \gamma_{\alpha k} \,,
\end{equation}
where $g_\alpha(t)$ interpolates smoothly between $0$ at $t=0$ and
$1$ within the switching time
$\tau_\text{sw}$.\cite{MuehlbacherUnpublished}($g_\alpha(t) =
    \sin^2[\pi t/(2\tau_\text{sw})]$ for $t < \tau_\text{sw}$ and
  $g_\alpha(t) = 1$ for $t \ge \tau_\text{sw}$)
In the limit of a quasi-continuous distribution of electronic
energies in the leads, the time-dependent contact transparency reads
\begin{eqnarray}
    \Gamma_\alpha(\epsilon,t) &=& 2\pi \rho_\alpha(\epsilon)|\gamma_{\alpha k}(t)|^2\equiv g_\alpha^2(t)\,
    \Gamma_\alpha(\epsilon) \, ,
\end{eqnarray}
where $\rho_\alpha(\epsilon)$ is the DOS of the leads.\cite{Jauho1994}
For the numerical implementation we choose a flat band with the width $2\epsilon_c$
and temperature-smoothed boundaries.\cite{Muhlbacher2008} The
corresponding profile of $\Gamma_\alpha(\epsilon)$ is given by
\begin{equation}
\label{eq:BandModel}
    \Gamma_\alpha(\epsilon) = \frac{\Gamma_\alpha}
    {\left[1 + e^{\beta(\epsilon - \epsilon_c)}\right]
    \left[1 + e^{-\beta(\epsilon + \epsilon_c)}\right]} \, .
\end{equation}

\begin{figure}
\begin{center}
    \includegraphics[scale=0.4]{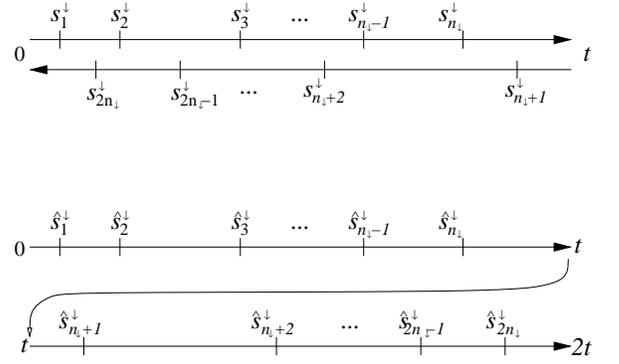}
    \caption{Contour-ordered (top) and time-ordered (bottom) sequences of tunneling times.
    \label{fig:contours}}
\end{center}
\end{figure}

Time-dependent transport properties can be calculated by a diagMC
scheme using the conventional breakdown of the time evolution
operator. Since the corresponding formalism has been described in
detail before,\cite{Werner2009,MuehlbacherUnpublished} we
restrict ourselves in the following to presenting the extensions
of the formalism, i.e.\ the inclusion of the probe electrode.

Expanding the time evolution operator in terms of the dot-lead
coupling yields a Dyson series where the integration variables
corresponds to the times at which tunneling between the dot and
the lead occurs. The contour-ordered kink sequence of the
$2\nt_\sigma$ tunneling times $s^\sigma_j$ of spin-$\sigma$
charges is denoted by
\begin{equation}
    \vec{s}^{\,\sigma}
    \equiv
    \{s^\sigma_1, s^\sigma_2, \dots, s^\sigma_{2\nt_\sigma}\} \,.
\end{equation}
They reside on the closed real-time contour $s: 0 \rightarrow t \rightarrow 0$
(c.f.\ Fig.\ \ref{fig:contours}). The corresponding time ordered analogon is denoted by
\begin{eqnarray}
    \hat{\vec{s}}_\sigma
    &\equiv&
    \{\hat{s}^\sigma_1, \hat{s}^\sigma_2, \dots, \hat{s}^\sigma_{2\nt_\sigma}\} \,,
\end{eqnarray}
with
\begin{eqnarray}
    \hat{s}^\sigma_j
    &=&
    \begin{cases}
    s^\sigma_j & \text{for}~j \le \n_\sigma \,,\\
    2t - s^\sigma_j & \text{for}~j > \n_\sigma \,,
    \end{cases}
\end{eqnarray}
where $\n_\sigma$ counts the number of spin-$\sigma$ tunneling
times on the forward time axis. Equation (\ref{eq:I_alpha}) can
then be conveniently expressed as
\begin{eqnarray}
\label{eq:I_alpha 2}
    I_\alpha(t)
    &=&
    -2e \sum_{\genfrac{}{}{0pt}{1}{\nt_\uparrow = 1}{\nt_\downarrow = 0}}^\infty
    (-1)^{\nt_\uparrow +\nt_\downarrow}
    \int_0^{2t}\!d\hat{\vec{s}}_\uparrow
    \int_0^{2t}\!d\hat{\vec{s}}_\downarrow
\nonumber\\
&&\times
    \text{Re}\!\left\{
    \mathcal{L}^{(\alpha)}_\uparrow(\hat{\vec{s}}_\uparrow)\,
    \mathcal{L}_\downarrow(\hat{\vec{s}}_\downarrow)\,
    \mathcal{G}(\hat{\vec{s}}_\uparrow,\hat{\vec{s}}_\downarrow)
    \right\} \,,
\end{eqnarray}
with the abbreviation
\begin{equation}
    \int_0^{2t}\!d\hat{\vec{s}}_\sigma
    \equiv
    \int_0^{2t}\!d\hat{s}^\sigma_{2\nt_\sigma}
    \int_0^{\hat{s}^\sigma_{2\nt_\sigma}}\!d\hat{s}^\sigma_{2\nt_\sigma - 1}
    \dots
    \int_0^{\hat{s}^\sigma_2}\!d\hat{s}^\sigma_1
\end{equation}
and the restriction that for $\sigma=\uparrow$ no integration is
performed over the fixed measurement time $\hat{s}_{n_\sigma+1} =
s_{n_\sigma+1} \equiv t$. The influence of the contacts is now
summarized in
\begin{eqnarray}
    \mathcal{L}_\downarrow(\hat{\vec{s}}_\downarrow)
    &=&
    (-1)^{\n_\downarrow} i^{\nt_\downarrow} \det\!\big(\Mat(\vec{s}_\downarrow)\big) \,,
\nonumber\\
    \mathcal{L}^{(\alpha)}_\uparrow(\hat{\vec{s}}_\uparrow)
    &=&
     i^{\nt_\uparrow} \det\!\big(\Mat^{(\alpha)}(\vec{s}_\uparrow)\big) \,,
\end{eqnarray}
where
\begin{eqnarray}
    \Mat_{ij}(\vec{s}_\downarrow)
    &=&
    \begin{cases}
    \hat{\Sigma}^{\lesser}(s^\downarrow_{2j-1}, s^\downarrow_{2i}) & \text{for}~i \le j \,,\\
    \hat{\Sigma}^{\greater}(s^\downarrow_{2j-1}, s^\downarrow_{2i}) & \text{for}~i > j \,,
    \end{cases}
\end{eqnarray}
and $\Mat^{(\alpha)}_{ij}(\vec{s}_\uparrow)$ is obtained from
$\Mat_{ij}(\vec{s}_\uparrow)$ by replacing $\hat{\Sigma}^{kl}$ by
$\hat{\Sigma}^{kl}_\alpha$ whenever one of the time
arguments is equal to the measurement time $s^\uparrow_{\n_\uparrow+1}=t$.
Here, $\hat{\Sigma}_\alpha^{\lesser}$ ($\hat{\Sigma}_\alpha^\greater$) denotes the
dot's lesser (greater) self energy with respect to the $\alpha$ contact, and
$\hat{\Sigma}^{kl} = \sum_\alpha \hat{\Sigma}_\alpha^{kl}$.
\footnote{Note that the selfenergy $\hat{\Sigma}$ of Sec.\ \ref{Sec:DQMC} is defined with
respect to the coupling to the leads $\Gamma$ while the selfenergy $\Sigma$ of Sec.\ \ref{Sec:PertTheory}
is defined with respect to the interaction strength $U$.}
Using the bandwidth profile (\ref{eq:BandModel}) we obtain
\begin{eqnarray}
    \lefteqn{
    \hat{\Sigma}_\alpha^{-k,k}(s,s')
    =
    -\frac{g(s)\, g(s')\Gamma_\alpha}{2\beta\sinh(\pi (s-s')/\beta)}
    \Bigg[
    }\quad
\nonumber\\
&&
    k\frac{e^{k\beta\mu_\alpha}}{1 - e^{-2\beta\epsilon_c}}
    \left(
    \frac{e^{-i\epsilon_c (s-s')}}
         {e^{k\beta\mu_\alpha} - e^{k\beta\epsilon_c}}
    - \frac{e^{i\epsilon_c (s-s')}}
         {e^{k\beta\mu_\alpha} - e^{-k\beta\epsilon_c}}
    \right)
\nonumber\\
&&{}+
    \frac{e^{-i\mu_\alpha (s-s')}}
        {\left(e^{-\beta\epsilon_c} - e^{-\beta\mu_\alpha}\right)
        \left(e^{-\beta\epsilon_c} - e^{\beta\mu_\alpha}\right)}
    \Bigg] \,,
\end{eqnarray}
with $k=\pm$.
Finally, the contribution from the dot operators to Eq.\ (\ref{eq:I_alpha 2}) is given by
\begin{equation}
    \mathcal{G}(\hat{\vec{s}}_\uparrow, \hat{\vec{s}}_\downarrow)
    =
    \mathcal{D}(\vec{s}_\uparrow) \,
    \mathcal{D}(\vec{s}_\downarrow) \,
    \mathcal{U}(\vec{s}_\uparrow, \vec{s}_\downarrow, \n_\downarrow) \,,
\end{equation}
with
\begin{equation}
    \mathcal{D}(\vec{s}_\sigma)
    =
    e^{i\epsilon_D (s^\sigma_1 - s^\sigma_2 + s^\sigma_3 - \dots)}
    e^{i\epsilon_D (s^\sigma_{2\nt_\sigma} - s^\sigma_{2\nt_\sigma-1} + s^\sigma_{2\nt_\sigma-2} - \dots)}
\end{equation}
and
\begin{equation}
    \mathcal{U}(\hat{\vec{s}}_\uparrow, \hat{\vec{s}}_\downarrow)
    =
    \exp\!\left\{iU\big[\tau^{(f)}_\text{double}(\vec{s}_\uparrow,
    \vec{s}_\downarrow) - \tau^{(b)}_\text{double}(\vec{s}_\uparrow,
    \vec{s}_\downarrow)\big]\right\} \,.
\end{equation}
Here $\tau^{(f/b)}_\text{double}$ denotes the time
during which the dot is doubly occupied on the forward/backward contour.

In order to calculate the conductance $g_M(t)$ as a function
of $\mu_M$ in the limit of vanishing tunneling coupling to the
third contact ($\Gamma_M \rightarrow 0$), we first note that
\begin{equation}
    \frac{d}{d\mu_M} \hat{\Sigma}^{-k,k}(s) = \frac{d}{d\mu_M}
    \hat{\Sigma}_M^{-k,k}(s)\,\propto\Gamma_M \,.
\end{equation}
Next we proceed by collecting all contributions to
\begin{equation}
    \frac{d}{d\mu_M} \det(\Mat)\, \det\!\big(\Mat^{(M)}\big)
\end{equation}
in lowest (linear) order in $\Gamma_M$ while neglecting all higher-order
terms.
In the limit $\Gamma_M\rightarrow0$ we find that
\begin{eqnarray}
    \frac{d}{d\mu_M}\det(\Mat)&\propto&\Gamma_M,
\\
    \det\!\big(\Mat^{(M)}\big)&\propto&\Gamma_M,
\\
    \frac{d}{d\mu_M}\det\!\big(\Mat^{(M)}\big)
    &=&
    \det\!\big(\tilde{\Mat}\big) +{\cal O}(\Gamma_M^{2})\,,
\end{eqnarray}
with
\begin{equation}
    \tilde{\Mat}_{ij}(\vec{s}_\uparrow) =
    \begin{cases}
    \frac{d}{d\mu_M}\hat{\Sigma}_M^\lesser(s^\uparrow_{2k-1}, t) &
    \text{for}~j =
    \n_\uparrow + 1,\, k \le j \,,\\
    \frac{d}{d\mu_M}\hat{\Sigma}_M^\greater(s^\uparrow_{2k-1}, t) & \text{for}~j = \n_\uparrow + 1,\, k > j \,,\\
    \hat{\Sigma}^\lesser(s^\uparrow_{2k-1}, s^\uparrow_{2j}) & \text{for}~j \neq \n_\uparrow+1,\, j \ge k \,,\\
    \hat{\Sigma}^\greater(s^\uparrow_{2k-1}, s^\uparrow_{2j}) & \text{for}~j
    \neq \n_\uparrow+1,\, j > k \,.
    \end{cases}
\end{equation}
Since $[d\det(\Mat)/d\mu_M] \det\!\big(\Mat^{(M)}\big)$ is quadratic in $\Gamma_M$
while $\det(\Mat) [d\det\!\big(\Mat^{(M)}\big)/d\mu_M]$ is linear, we arrive at
\begin{equation}
    \frac{d}{d\mu_M}
    \det\!\big(\Mat^{(M)}\big)
    \det\!\big(\Mat\big)
    =
    \det\!\big(\tilde{\Mat}\big)\,
    \det\!\big(\Mat\big) +{\cal O}(\Gamma_M^2),
\end{equation}
and the third-terminal conductance is found to read
\begin{eqnarray}
\label{g_M}
    \lefteqn{
    \lim_{\Gamma_M \rightarrow 0} \frac{g_M(t)}{\Gamma_M}
    =
    }
\nonumber\\
&&
    -2e
    \sum_{\genfrac{}{}{0pt}{1}{\nt_\uparrow = 1}{\nt_\downarrow =0}}^\infty
    (-1)^{\nt_\uparrow +\nt_\downarrow}
    \int_0^{2t}\!d\hat{\vec{s}}_\uparrow
    \int_0^{2t}\!d\hat{\vec{s}}_\downarrow
    (-1)^{\n_\downarrow}
\nonumber\\
&&\quad\times
    \text{Re}\Bigg\{
    \frac{\det\!\big(\tilde{\Mat}(\vec{s}_\uparrow)\big)}{\Gamma_M}
    \det\!\big(\Mat(\vec{s}_\downarrow)\big)\,
    \mathcal{G}(\hat{\vec{s}}_\uparrow, \hat{\vec{s}}_\downarrow)
    \Bigg\} \,.
\end{eqnarray}
Note that in the zero temperature limit,
\begin{equation}
    \frac{d}{d\mu_M} \hat{\Sigma}_\alpha^{-k,k}(s,s')
    \stackrel{T=0}{=}
    i\frac{g(s)g(s')\Gamma_\alpha}{2\pi}
    e^{-i\mu_\alpha (s-s')}
\end{equation}
so that $g_M(t)$ becomes independent of the bandwidth of the third contact.

\begin{figure}
\begin{center}
   \includegraphics[width=\columnwidth]{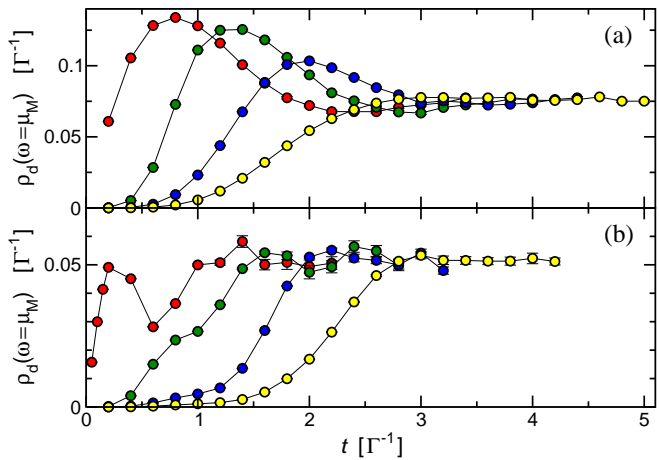}
   \caption{Spectral function $\rho_d(\omega=\mu_M)$ at zero temperature and $V = 2\Gamma$ for
   (a) $U=0=\epsilon_D$, $\mu_M = 2\Gamma$ and  (b) $U =4\Gamma=-\epsilon_D/2$,
   $\mu_M=4\Gamma$.  Both panels show data corresponding to four different switching times
   $\tau_\text{sw} = 0$, $\Gamma^{-1}$, $2\Gamma^{-1}$, and $3\Gamma^{-1}$
   (red, green, blue, and yellow, respectively). The lines are guides to the eye only.
   \label{fig:SteadyState}}
\end{center}
\end{figure}

A crucial point in the simulation is the attainment of the steady
state, in which the transport current stops to be time-dependent.
The further the system is propagated in time, starting from the
initial time $t=0$ when the tunneling is switched on, the closer
the measured current is to the actual steady state value.
Unfortunately, this time evolution requires a rapidly increasing
computation time. Still it has been shown in recent
works\cite{Muhlbacher2008,Werner2010,MuehlbacherUnpublished} that
in many cases the steady state is indeed accessible with moderate
numerical effort. As shown in Fig.\ \ref{fig:SteadyState}, in our
case the steady state of $g_M(t)$ is typically preceded by
strongly non-monotonic dynamics. These elongate the transient
regime and increase the timescale on which the steady-state regime
is reached. However, this timescale can be vastly reduced by
adopting a smooth switching of the tunneling coupling according to
Eq.~(\ref{t(t)}), thereby extending
significantly the parameter regime for which the steady state can
be reached.\cite{MuehlbacherUnpublished}

\begin{figure}
\begin{center}
   \includegraphics[width=\columnwidth]{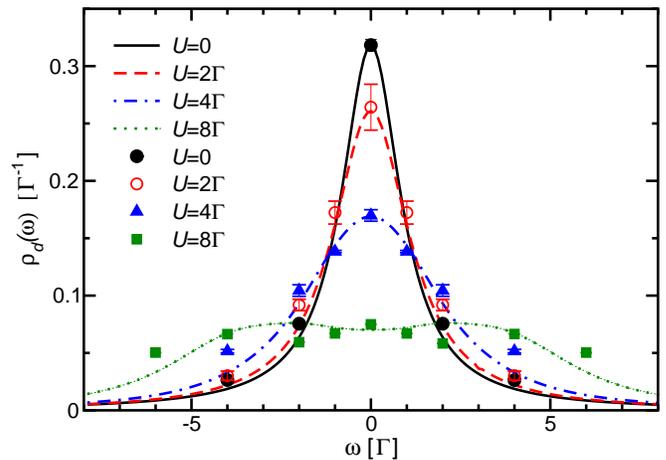}
   \caption{Zero temperature dot spectral function for different
   interaction strengths and fixed bias voltage $V/\Gamma=2.$
   The points are numerical diagMC data and curves are the analytical
   results in the second order in $U$ perturbation theory. The
   slight mismatch of the $U=0$ data points and the analytical curve
   is due to finite bandwidth used in the numerical simulations (i.~e.
   a
     bandwidth of $2\epsilon_c/\Gamma
     = 12$ for $U/\Gamma \le 4$ and $2\epsilon_c/\Gamma
     = 20$ for $U/\Gamma = 8$).
   \label{fig:CompareData:DQMC}}
\end{center}
\end{figure}

\section{Results and discussion}
\label{Sec:Discussion} We would like to summarize the results based on
the second order self-energy first, see Fig.\ \ref{fig:Results:PertTheory}. This
approximation is known to \emph{qualitatively} reflect all
features one expects in the low energy sector. At vanishing
voltage one obtains the typical trident-shaped curve. The central
peak in this particular case can be regarded as the precursor of
the in reality much sharper Kondo (Abrikosov-Suhl)
resonance\cite{Hewson1997} having a width (in equilibrium) which can
be estimated as\cite{Tsvelick1983}
\begin{eqnarray}
    \label{eq:TK}
    T_K \sim \frac{\sqrt{2 U \Gamma}}{\pi} e^{-\pi U/(8\Gamma)} \, .
\end{eqnarray}
This formula gives $T_K \sim 0.055 \Gamma$ for $U/\Gamma=8$, which
is much smaller than the actual width of the resonance
$\widetilde{T}_K \sim \Gamma$. The two much wider peaks
(shoulders), which are located at $\omega \approx \pm U/2$ are the
Hubbard sidebands. Keeping $T=0$ and increasing the bias voltage
does not appear to produce any noticeable qualitative change until
$V$ hits the threshold of $\approx \Gamma$, beyond which the
`Kondo'-peak rapidly deteriorates and completely disappears for $V
> \Gamma$. Interestingly, a very similar destruction of the central
peak can be observed in equilibrium $V=0$ at finite
temperature $T>\tilde{T}_K$. We thus conjecture that the
effect of the finite voltage might be captured by an effective
temperature $T_{\rm eff} \sim V$.

Further insight is gained through the numerical
\mbox{diagMC} simulations for the spectral function at finite voltage and
$U$. The numerical data, shown in Fig.\
\ref{fig:CompareData:DQMC}, turn out to be in an excellent agreement with
the analytical results using the second order self-energy up to
the interaction strength $U/\Gamma = 2$. The matching of the perturbative curve
and numerical data is still agreeable even at $U/\Gamma = 4$. At
$U/\Gamma = 8$ one recognizes the very weak remnants of the Kondo
peak at $\omega=0$ while the rest of the curve has a similar
qualitative behavior as the perturbative result and the same order
of magnitude.
This suggests that the complicated collective multi-particle effects
contained in the full self-energy (i.e.\ exact in $U$) only have a weak effect
and that the main information is already accounted for by the
lowest order self-energy discussed in Sec.~\ref{Sec:PertTheory}.
This observation is again compatible with the existence of an effective temperature.
We would like to emphasize that $T_{\rm eff}$ is different from the decoherence rate discussed, e.~g., in Refs.\
[\onlinecite{Kaminski2000,Rosch2001,Kehrein2005,Schoeller2009,Pletyukhov2010}] as this is not defined in the perturbative regime.

In the voltage regimes considered, diagMC does not produce any
evidence for the Kondo peak splitting. The relatively high applied
voltage $V \sim T_{\rm eff} \gg T_K$ appears to induce a widening
of the Kondo resonance, which renders the observation of the peak
splitting impossible.
For the diagMC approach outlined in Sec.~\ref{Sec:DQMC}, smaller
voltages lead to an increase of the timescale over which time-dependent
transport properties have to be monitored until a stable stationary regime
can be identified. This significantly increases the computational effort
necessary to extract the steady-state values.
Therefore, in the system under consideration and the numerical scheme and equipment used, it is not yet possible to
give a final answer to the question of the Kondo peak doubling with satisfactorily precision.



A very interesting issue is the characteristic time scale
necessary for the Kondo effect to fully develop. In the context of
a sudden gate voltage quench this problem was discussed in Ref.\
[\onlinecite{Nordlander1999}]. On general grounds one would expect
that the Kondo peak develops on the time scale $\sim 1/T_K$, as
$T_K$ is the only energy scale available in the system in
equilibrium. We observe, however, that even in the case of large
$U = 8 \Gamma$ the steady state of the spectral density at the
position of the Kondo peak $\omega=0$ does not take a longer time to
establish than for the energies outside of the Kondo peak. The
only exception are the Hubbard subbands $\omega \approx \pm U/2$,
where the time development appears to be very slow.

To summarize, we present numerical and analytical results for the
impurity spectral function of the symmetric Anderson model in
nonequilibrium at zero temperature. We find an excellent agreement
of the numerically exact diagMC data with the perturbative
expansion of the lowest order in interaction. For larger $U$ we
observe a small peak due to the Kondo resonance, which does not
show a splitting due to finite applied voltage.

\acknowledgements The authors are supported by the Kompetenz\-netz
``Funktionelle Nanostrukturen III'' of the Baden-W\"urttemberg
Stiftung (Germany) and by the DFG under grant No. KO 2235/3. L.M.
acknowledges the use of the computing resources provided by the
Black Forest Grid Initiative. The authors would like to thank
A.~O.~Gogolin and H.~Grabert for many interesting discussions.

\bibliography{MC_Anderson_Kondo}

\end{document}